\documentclass[8pt, colorlinks, a4paper]{article}
\usepackage[utf8]{inputenc}
\usepackage[T1]{fontenc}
\usepackage{fixltx2e}
\usepackage{graphicx}
\usepackage{longtable}
\usepackage{float}
\usepackage{wrapfig}
\usepackage{rotating}
\usepackage[normalem]{ulem}
\usepackage{amsmath}
\usepackage{textcomp}
\usepackage{marvosym}
\usepackage{wasysym}
\usepackage{amssymb}
\usepackage{hyperref}
\usepackage{graphicx}
\usepackage[utf8]{inputenc}
\usepackage[T1]{fontenc}
\usepackage{lmodern}
\usepackage{amsmath}
\usepackage{microtype} % Slightly tweak font spacing for aesthetics
\usepackage{geometry}
\hypersetup{allcolors = mycolor}
\usepackage{setspace, caption}
\renewcommand\ref{}

\usepackage{lineno}
\usepackage[figuresonly, nolists]{endfloat}
\usepackage{biology_citations}
\usepackage{color}

\definecolor{mycolor}{rgb}{0.2,0.2,0.8}

\renewcommand{\url}[1]{\underline{\textcolor{blue}{#1}}}

\author{Yann Zerlaut and Alain Destexhe}

\date{\today}

\title{A mean-field model for conductance-based networks of adaptive
  exponential integrate-and-fire models}

\begin{document}

\maketitle

%\linenumbers

\section{Abstract}
\label{sec-1}
\bfseries

Voltage-sensitive dye imaging (VSDi) has revealed fundamental
properties of neocortical processing at mesoscopic scales.  Since VSDi
signals report the average membrane potential, it seems natural to use
a mean-field formalism to model such signals.  Here, we investigate a
mean-field model of networks of Adaptive Exponential (AdEx)
integrate-and-fire neurons, with conductance-based synaptic
interactions.  The AdEx model can capture the spiking response of
different cell types, such as regular-spiking (RS) excitatory neurons
and fast-spiking (FS) inhibitory neurons.  We use a Master Equation
formalism, together with a semi-analytic approach to the transfer
function of AdEx neurons.  We compare the predictions of this
mean-field model to simulated networks of RS-FS cells, first at the
level of the spontaneous activity of the network, which is well
predicted by the mean-field model.  Second, we investigate the
response of the network to time-varying external input, and show that
the mean-field model accurately predicts the response time course of
the population.  One notable exception was that the ``tail'' of the
response at long times was not well predicted, because the mean-field
does not include adaptation mechanisms.  We conclude that the Master
Equation formalism can yield mean-field models that predict well the
behavior of nonlinear networks with conductance-based interactions and
various electrophysiolgical properties, and should be a good candidate
to model VSDi signals where both excitatory and inhibitory neurons
contribute.

\normalfont

\section{Introduction}
\label{sec-2}
\normalfont

Recent advances in imaging technique, in particular voltage-sensitive
dye imaging (VSDi), have revealed fundamental properties of
neocortical processing
\cite{Arieli1996,Contreras2001,Petersen2001,Ferezou2006,Civillico2012}:
subthreshold responses to sensory inputs are locally homogeneous in
primary sensory areas, depolarizations tend to spread across spatially
neighboring regions and responses to sensory stimuli are strongly
affected by the level of ongoing activity. It also appears as a great
tool to unveil how the spatio-temporal dynamics in the neocortex shape
canonical cortical operations such as normalization
\cite{Reynaud2012}.

On the other hand, the literature lacks, to the best of our knowledge,
theoretical models that provides a detailed account of those phenomena
with a clear relation between the biophysical source of the VSDi
signal and network dynamics at that spatial scale (i.e. at the
millimeters or centimeters scale). Detailed model of a neocortical
column (i.e. \(\sim\)0.5mm$^{\text{2}}$ scale) have been recently
proposed, see \citetext{Chemla2010} for the link with the VSDi signal
or more generally \citetext{Markram2015}, but their computational cost
impedes the generalization to higher spatial scale. The aim of the
present communication is therefore to design a theoretical model of
neocortical dynamics with the following properties: 1) it should
describe the temporal scale of optical imaging as well as easily
extend to its spatial scale and 2) it should have a correlate in terms
of single-cell dynamics (in particular membrane potential dynamics),
so that the model can directly generate predictions for the signal
imaged by the VSDi technique \cite{Berger2007}.

More specifically, our study focuses on network dynamics in
\emph{activated} cortical states, thus the desired model should
describe neocortical computation in the asynchronous regime, where
cortical activity is characterized by irregular firing and strong
subthreshold fluctuations at the neuronal level
\cite{Steriade2001,Destexhe2003}. The strategy behind the present
model is to take advantage of the \emph{mean-field} descriptions of
network dynamics in this regime. Via self-consistent approaches, those
descriptions allow to capture the dynamical properties of population
activity in recurrent networks
\cite{Amit1997,Brunel1999,Brunel2000,Latham2000,ElBoustani2009}. The
present model will thus consider randomly connected network of 10000
neurons as a unit to describe a cortical column and we will compare
its behavior to network simulations.

\section{Material and Methods}
\label{sec-3}
\small

We describe the equations and parameters used for the neuronal,
synaptic and network modeling. We present our \emph{heuristic}
treatment of the neuronal \emph{transfer functions}: the quantity that
accounts for the cellular computation in \emph{mean-field} models of
population activity. Then, we present the specific markovian model of
population activity used in this study. Finally, we compare this
analytical model to numerical simulations of network dynamics.

\subsection{Single neuron models}
\label{sec-3-1}

The neuronal model used in this study is the adaptative exponential
and fire (AdExp) model \cite{Brette2005a}. The equation for the membrane
potential and the adaptation current therefore reads:

\begin{equation}
\label{eq:iAdExp}
  \left\{
  \begin{split}
  & C_m\,\frac{dV}{dt} = g_{L} \,(E_{L}-V) + I_{syn}(V,t) + k_a e^{\frac{V - V_{thre} }{k_a}}- I_w \\
  & \tau_w \frac{d I_w}{dt} = - I_w + \sum_{t_s \in \{t_{spike}\}} b \, \, \delta (t-t_s)
  \end{split}
\right.
\end{equation}

where $I_{syn}(V, t)$ is the current emulating synaptic activity that
will create the fluctuations, $I_w$ reproduces the I$_{\text{m}}$ current
\cite{McCormick1985}. The spiking mechanism is the following: when
$V(t)$ reaches \(V_{thre}+5 \, k_a \), this triggers a spike t$_{\text{s}}$ $\in$
\{t$_{\text{spike}}$\}, this increases the adaptation variable $I_w$ by \(b\),
the membrane potential is then clamped at \(E_L\) for a duration
$\tau$$_{\text{refrac}}$=5ms. We consider two versions of this model: a regular
spiking neuron for the excitatory cells and a fast spiking neuron for
the inhibitory cells (see Figure \ref{fig:tf}). The parameters of those two
models can be found on Table \ref{table:params}.

\begin{table*}[tb!]
\caption{\label{table:params}\textbf{Model parameters}.}
\centering
\begin{tabular}{l|l|l|r|l}
\textbf{Parameters} & Parameter Name & Symbol & Value & Unit\\
\hline
 &  &  &  & \\
\textbf{cellular properties} &  &  &  & \\
 & leak conductance & \(g_L\) & 10 & nS\\
 & leak reversal potential & \(E_L\) & -65 & mV\\
 & membrane capacitance & \(C_m\) & 150 & pF\\
 & leak reversal potential & \(E_L\) & -65 & mV\\
 & AP threshold & \(V_{thre}\) & -50 & mV\\
 & refractory period & \(\tau_{refrec}\) & 5 & ms\\
 & adaptation time constant & \(\tau_w\) & 500 & ms\\
\textbf{excitatory cell} &  &  &  & \\
 & sodium sharpness & \(k_a\) & 2 & mV\\
 & adaptation current increment & \(b\) & 20 & pA\\
 & adaptation conductance & \(a\) & 4 & nS\\
\textbf{inhibitory cell} &  &  &  & \\
 & sodium sharpness & \(k_a\) & 0.5 & mV\\
 & adaptation current increment & \(b\) & 0 & pA\\
 & adaptation conductance & \(a\) & 0 & nS\\
\textbf{synaptic properties} &  &  &  & \\
 & excitatory reversal potential & \(E_e\) & 0 & mV\\
 & inhibitory reversal potential & \(E_i\) & -80 & mV\\
 & excitatory quantal conductance & \(Q_e\) & 1 & nS\\
 & inhibitory quantal conductance & \(Q_i\) & 5 & nS\\
 & excitatory decay & \(\tau_e\) & 5 & ms\\
 & inhibitory  decay & \(\tau_i\) & 5 & ms\\
\textbf{numerical network} &  &  &  & \\
 & cell number & \(N_{tot}\) & 10000 & \\
 & connectivity probability & \(\epsilon\) & 5\% & \\
 & fraction of inhibitory cells & g & 20\% & \\
 & external drive & \(\nu_e^{drive}\) & 4 & Hz\\
\textbf{ring model} &  &  &  & \\
 & total extent & \(L_{tot}\) & 40 & mm\\
 & excitatory connectivity extent & \(l_{exc}\) & 5 & mm\\
 & inhibitory connectivity extent & \(l_{inh}\) & 1 & mm\\
 & propagation delay & \(v_c\) & 300 & mm/s\\
\end{tabular}
\end{table*}

\subsection{Synaptic model}
\label{sec-3-2}

The time- and voltage-dependent current that stimulate the neuron is
made of the sum of an excitatory and inhibitory currents (indexed by
\(s \in \{e,i\}\) and having a reversal potential \(E_s\)):

\begin{equation}
\label{eq:syn-current}
 I_{syn}(V,t) = \sum_{s \in \{e,i\}} \sum_{t_s \in \{t_s\}} Q_s \, e^{-\frac{t}{\tau_s}} \, (E_{s}-V) \, \mathcal{H}(t-t_s)
\end{equation}
where \(\mathcal{H}\) is the Heaviside function.

This synaptic model is referred to as the \emph{conductance-based
  exponential} synapse. The set of events \(\{t_e\}\) and \(\{t_i\}\)
are the set of excitatory and inhibitory events arriving to the
neuron. In numerical simulations of single neurons (performed to
determine the \textit{transfer function} \(\mathcal{F}\) of either
excitatory or inhibitory neurons), it will be generated by stationary
Poisson processes. On the other hand, in numerical simulations of
network dynamics it will correspond to the set of spike times of the
neurons connecting to the target neurons, both via recurrent and
feedforward connectivity.

\subsection{Numerical network model}
\label{sec-3-3}

\begin{figure}[tb!]
\centering
\includegraphics[width=.5\linewidth]{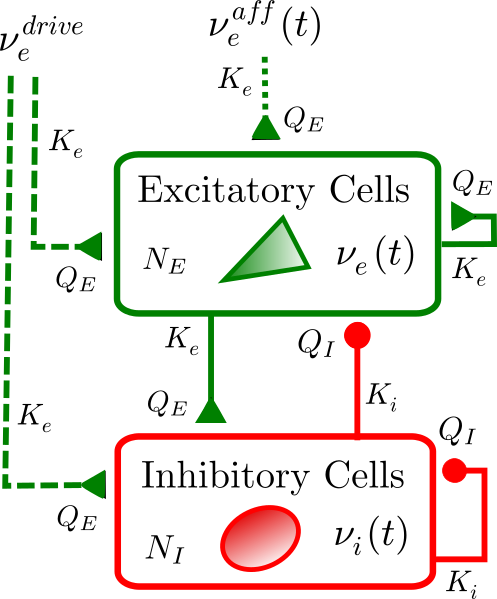}
\caption{\label{ntwk-architect}\textbf{Schematic of the local network architecture}. The network is made of \(N_e=(1-g) \, N_{tot}\) excitatory and \(N_i=g \, N_{tot}\) inhibitory neurons. All excitatory connections (afferent and recurrent) onto a neuron corresponds to \(K_e= \epsilon \, (1-g) \, N_{tot}\) synapses of weight \(Q_e\). All inhibitory connections (afferent and recurrent) onto a neuron corresponds to \(K_i= \epsilon \, g \, N_{tot}\) synapses of weight \(Q_i\)}
\end{figure}

All simulations of numerical network were performed with the \texttt{brian2}
simulator \cite{Goodman2009}, see \url{http://brian2.readthedocs.org}. For all
simulations, the network was composed of \(N_{tot}\)=10000 neurons,
separated in two populations, one excitatory and one inhibitory with a
ratio of g=20\% inhibitory cells. Those two populations we recurrently
connected (internally and mutually) with a connectivity probability
\(\epsilon\)=5\%.

Because this network did not display self-sustained activity (see
Figure \ref{fig:phase-space}, in contrast to \citetext{Vogels2005}),
an excitatory population exerted an \emph{external drive} to bring the
network out of the quiescent state. This population targeted both the
excitatory and inhibitory neurons. Note that the firing rate of this
population was linearly increased to avoid a too strong initial
synchronization (see Figure \ref{fig:ntwk-act}). Finally, when
studying responses to external inputs, an excitatory population of
time varying firing rate was added to evoke activity transients in the
population dynamics. This last stimulation targeted only the
excitatory population. The number of neurons in those two excitatory
populations was taken as identical to the number of excitatory neurons
(i.e. \((1-g)\,N_{tot}\)) and created synapses onto the recurrent
network with the same probability \(\epsilon\). After temporal
discretization, the firing rates of those afferent populations were
converted into spikes by using the properties of a Poisson process
(i.e. eliciting a spike at \(t\) with a probability \(\nu(t) \,dt
\)). All simulations were performed with a time-step dt=0.1ms.

\subsection{Estimating the transfer functions of single neurons}
\label{sec-3-4}

The transfer function \(\mathcal{F}\) of a single neuron is defined
here as the function that maps the value of the stationary excitatory
and inhibitory presynaptic release frequencies to the output
stationary firing rate response, i.e. \(\nu_{out} = \mathcal{F}(\nu_e,
\nu_i)\). Note the stationary hypothesis in the definition of the
transfer function (see discussion in main text).

Because an analytical solution of this function for the single neuron
models considered in our study is a very challenging mathematical
problem, we adopted a semi-analytical approach. We performed numerical
simulations of single cell dynamics at various excitatory and
inhibitory presynaptic frequencies (\(\nu_e\) and \(\nu_i\)
respectively) (see the output in Figure \ref{fig:tf}) on which we fitted the
coefficients of an analytical template to capture the single cell
model's response.

The procedure relied on fitting a \emph{phenomenological threshold}
\(V_{thre}^{eff}\) that accounts for the single neuron non-linearities
(spiking and reset mechanism, adaptation mechanisms) on top of the
subthreshold integration effects \cite{Zerlaut2016}. This
phenomenological threshold is then plugged-in into the following
formula (analogous to \citetext{Amit1997}) to become our firing response
estimate:

\begin{equation}
\label{eq:template}
  \nu_{out} = \frac{1}{2\,\tau_V} \cdot 
  {Erfc}(\frac{V_{thre}^{eff}-\mu_V}{\sqrt{2} \, \sigma_V})
\end{equation}

Where \( (\mu_V, \sigma_V, \tau_V) \) are the mean, standard deviation
and autocorrelation time constant of the membrane potential
fluctuations. How to calculate those quantities as a response to a
stationary stimulation is the focus of the next section.

The phenomenological threshold was taken as a second order polynomial
in the three dimensional space \( (\mu_V, \sigma_V, \tau_V) \):

\begin{equation}
\label{eq:quadratic-threshold}
  \begin{split}
   V_{thre}^{eff} & (\mu_V, \sigma_V, \tau_V^N) = P_0 + 
   \sum_{x \in \{ \mu_V, \sigma_V, \tau_V^N\}} P_x \cdot \Big(  \frac{x - x^0}{\delta x^0} \Big) + \\
   & \sum_{ x,y \in \{ \mu_V, \sigma_V, \tau_V^N\}^2} P_{xy} \cdot
   \Big( \frac{x - x^0}{\delta x^0}  \Big) \,  \Big( \frac{y - y^0}{\delta y^0} \Big)
 \end{split}
\end{equation}

Where the normalization factors \(\mu_V^0\)=-60mV, \(\delta
\mu_V^0\)=10mV, \(\sigma_V^0\)=4mV, \(\delta \sigma_V^0\) = 6mV,
\(\tau_V^{N0}\)=0.5 and \(\delta \tau_V^{N0}\)= 1 arbitrarily delimits
the \emph{fluctuation-driven} regime (a mean value $x$ and an extent
$\delta x$, \(\forall x \in \{\mu_V, \sigma_V, \tau_V^N\}\)).  They
render the fitting of the phenomenological threshold easier, as they
insure that the coefficients take similar values. It is kept constant
all along the study. The phenomenological threshold was taken as a
second order polynomial and not as a linear threshold, for two
reasons: 1) unlike in an experimental study \cite{Zerlaut2016}, we are
not limited by the number of sampling points, the number of fitted
coefficients can thus be higher as the probability of overfitting
becomes negligible 2) it gives more flexibility to the template,
indeed the linear threshold was found a good approximation in the
\emph{fluctuation-driven} regime, i.e. when the diffusion approximation
holds, however, for low values of the presynaptic frequencies, we can
be far from this approximation, the additional coefficients are used
to capture the firing response in those domains.

The fitting procedure was identical to \citetext{Zerlaut2016}, it
consisted first in a linear regression in the phenomenological
threshold space of Equation \ref{eq:quadratic-threshold}, followed by a
non-linear optimization of Equation \ref{eq:template} on the firing rate
response. Both fitting were performed with the \texttt{leastsq} method in the
\texttt{optimize} package of \texttt{SciPy}.

\subsection{Calculus of the subthreshold membrane potential fluctuations}
\label{sec-3-5}

Here, we detail the analytical calculus that translate the input to
the neuron into the properties of the membrane potential
fluctuations. The input is made of two Poisson shotnoise: one
excitatory and one inhibitory that are both convoluted with an
exponential waveform to produce the synaptic conductances time
courses.

\subsubsection{Conductances fluctuations}
\label{sec-3-5-1}

From Campbell's theorem \cite{Papoulis1991}, we first get the mean
(\(\mu_{Ge}, \mu_{Gi}\)) and standard deviation (\(\sigma_{Ge},
\sigma_{Gi}\)) of the excitatory and inhibitory conductance
fluctuations:

\begin{equation}
\begin{split}
& \mu_{Ge}(\nu_e, \nu_i) = \nu_e \, K_e \, \tau_e \, Q_e \\
& \sigma_{Ge}(\nu_e, \nu_i) = \sqrt{\frac{\nu_e \, K_e \, \tau_e}{2}} \, Q_e \\
& \mu_{Gi}(\nu_e, \nu_i) = \nu_i \, K_i \, \tau_i \, Q_i \\
& \sigma_{Gi}(\nu_e, \nu_i) = \sqrt{\frac{\nu_i \, K_i \, \tau_i}{2}} \, Q_i 
\end{split}
\end{equation}

The mean conductances will control the input conductance of the neuron
\(\mu_G\) and therefore its effective membrane time constant
\(\tau_m\):

\begin{equation}
\begin{split}
& \mu_{G}(\nu_e, \nu_i) = \mu_{Ge} + \mu_{Gi} + g_L \\
& \tau_m(\nu_e, \nu_i) = \frac{C_m}{\mu_{G}}
\end{split}
\end{equation}

\subsubsection{Mean membrane potential}
\label{sec-3-5-2}

Following \citetext{Kuhn2004}, the mean membrane potential is obtained
by taking the stationary solution to static conductances given by the
mean synaptic bombardment (for the passive version of Equation
\ref{eq:iAdExp}, i.e. removing the adaptation and spiking mechanisms). We
obtain:

\begin{equation}
\label{eq:mu-v}
\mu_V(\nu_e, \nu_i) = \frac{\mu_{Ge} \, E_e + \mu_{Gi} \, E_i + g_L \, E_L}{\mu_{G}}
\end{equation}

We will now approximate the driving force \(E_s - V(t)\) of synaptic
events by the level resulting from the mean conductance bombardment:
\(E_s - \mu_V\). This will enable an analytical solution for the
standard deviation \( \sigma_V\) and the autocorrelation time \(
\sigma_V\) of the fluctuations.

\subsubsection{Power spectrum of the membrane potential fluctuations}
\label{sec-3-5-3}

Obtaining \( \sigma_V\) and \(\tau_V\) is achieved by computing the
power spectrum density of the fluctuations. In the case of Poisson
processes, the power spectrum density of the fluctuations resulting
from the sum of events \(PSP_{s}(t)\) at frequency \(K_{s} \, \nu_s\)
can be obtained from shotnoise theory \cite{Daley2007}:

\begin{equation}
 P_V(f)  = \sum_{s \in \{e,i\}} K_s \, \nu_{s} \, \| \hat{\mathrm{PSP}_s}(f) \|^2
\end{equation}

where \(\hat{\mathrm{PSP}_s}(f)\) is the Fourier transform of the
time-varying function \(\mathrm{PSP}(t)\). Note that the relations
presented in this paper rely on the following convention for the
Fourier transform: \( \hat{F}(f) = \int_\mathbb{R} F(t) \, e^{- 2 i
\pi f t} \, dt\).

After fixing the driving force to \(E_s - \mu_V\), the equation for a
 post-synaptic membrane potential event \(s\) around \(\mu_V\) is

\begin{equation}
 \tau_m \frac{d \, \mathrm{PSP}_s }{dt} + \mathrm{PSP}_s = U_s \, \mathcal{H}(t) \, e^{\frac{-t}{\tau_s}}
\end{equation}

where \( U_s = \frac{Q_s}{\mu_G} (E_s - \mu_V) \) and \(
\mathcal{H}(t) \) is the Heaviside function.

Its solution is:

\begin{equation}
\mathrm{PSP}_s(t)  = U_s \, \frac{\tau_s}{\tau_m - \tau_s} \, \big( 
e^{\frac{-t}{\tau_m}} - e^{\frac{-t}{\tau_s}} \big) \, \mathcal{H}(t)
\end{equation}

We take the Fourier transform:

\begin{equation}
\hat{\mathrm{PSP}_s}(f) = U_s \, \frac{\tau_s}{\tau_m - \tau_s} \, 
\big(
\frac{\tau_{m}}{2 \, i  \,  \pi \, f \, \tau_{m} +1} 
- \frac{\tau_s}{2 \, i  \,  \pi \, f \, \tau_s +1} \big)
\end{equation}

We will need the value of the square modulus at \(f=0\):

\begin{equation}
\label{eq:psp0}
\| \hat{\mathrm{PSP}}(0) \|^2 = (U_s \cdot \tau_s)^2
\end{equation}

As well as the integral of the square modulus:

\begin{equation}
\label{eq:psp-int}
\int_\mathbb{R}  df \, \| \hat{\mathrm{PSP}}(f) \|^2 = \frac{(U_s \cdot \tau_s)^2}{2 \, (\tau_\mathrm{m}^\mathrm{eff} + \tau_s ) }
\end{equation}

\subsubsection{Standard deviation of the fluctuations}
\label{sec-3-5-4}

The standard deviation follows:

\begin{equation}
 (\sigma_V)^2  = \int_\mathbb{R}  df \, P_V(f)
\end{equation}

Using Equation \ref{eq:psp-int}, we find the final expression for \(\sigma_V\):

\begin{equation}
\label{eq:sigma-v}
 \sigma_V(\nu_e, \nu_i)  = \sqrt{ \sum_s K_s \, \nu_s \, \frac{(U_s \cdot \tau_s)^2}{2 \, (\tau_\mathrm{m}^\mathrm{eff} + \tau_s ) } }
\end{equation}

\subsubsection{Autocorrelation-time of the fluctuations}
\label{sec-3-5-5}

We defined the global autocorrelation time as \cite{Zerlaut2016}:

\begin{equation}
  \tau_V = \frac{1}{2} \, \big( \frac{\int_\mathbb{R} P_V(f) \, d f}{ P_V(0) } \big)^{-1}
\end{equation}

Using Equations \ref{eq:psp-int} and \ref{eq:psp0}, we find the final expression
for \(\tau_V\):

\begin{equation}
\label{eq:tau-v}
  \tau_V(\nu_e, \nu_i) = \Big( \frac{
  \sum_s \big( K_s \, \nu_s \, (U_s \cdot \tau_s)^2\big) 
  }{
  \sum_s \big( K_s \, \nu_s \, (U_s \cdot \tau_s)^2 /(\tau_\mathrm{m}^\mathrm{eff} + \tau_s ) \big)
  } \Big)
\end{equation}

Therefore the set of Equations \ref{eq:mu-v}, \ref{eq:sigma-v} and \ref{eq:tau-v}
translate the presynaptic frequencies into membrane fluctuations
properties \(\mu_V, \sigma_V, \tau_V\).

The previous methodological section allowed to translate the
fluctuations properties \(\mu_V, \sigma_V, \tau_V\) into a spiking
probability thanks to a minimization procedure. The combination of the
present analytical calculus and the previous fitting procedure (on
numerical simulations data) constitute our semi-analytical approach to
determine the transfer function of a single cell model: \( \nu_{out} =
\mathcal{F}(\nu_e, \nu_i)\).

\subsection{Master equation for local population dynamics}
\label{sec-3-6}

An analytical description of the cellular transfer function is the
core of theoretical descriptions of asynchronous dynamics in sparsely
connected random networks \cite{Amit1997,Brunel2000,Renart2004}.

Because we will investigate relatively slow dynamics
(\(\tau\)>25-50ms) (and because of the stationary formulation of our
transfer function), we will use the Markovian description developed
in \citetext{ElBoustani2009}, it describes network activity at a time
scale \(T\), for which the network dynamics should be Markovian. The
choice of the time-scale \(T\) is quite crucial in this formalism, it
should be large enough so that activity can be considered as
memoryless (e.g. it can not be much smaller than the refractory
period, that would introduce memory effects) and small enough so that
each neuron can fire statistically only once per time interval
\(T\). Following \citetext{ElBoustani2009}, we will arbitrarily take
\(T\)=5ms all along the study as it offers a good compromise between
those two constraints.

The formalism describes the first and second moments of the population
activity for each populations. We consider here two populations: one
excitatory and one inhibitory, the formalism thus describes the
evolution of five quantities: the two means \(\nu_e(t)\) and
\(\nu_i(t)\) of the excitatory and inhibitory population activity
respectively (the instantaneous population firing rate, i.e. after
binning in bins of \(T\)=5ms, see discussion in
\citetext{ElBoustani2009}), the two variances \(c_{ee}(t)\) and
\(c_{ii}(t)\) of the the excitatory and inhibitory population activity
respectively and the covariance \(c_{ei}(t)\) between the excitatory
and inhibitory population activities. The set of differential
equations followed by those quantities reads \cite{ElBoustani2009}:

\begin{equation}
\label{eq:master-equation}
\left\{
\begin{split}
T \, \frac{\partial \nu_\mu}{\partial t} = & (\mathcal{F}_\mu - \nu_\mu )
   + \frac{1}{2} \, c_{\lambda \eta} \, 
\frac{\partial^2 \mathcal{F}_\mu}{\partial \nu_\lambda \partial \nu_\eta} \\
T \, \frac{\partial c_{\lambda \eta} }{\partial t}  =  & A_{\lambda \eta} +
(\mathcal{F}_\lambda - \nu_\lambda ) \, (\mathcal{F}_\eta - \nu_\eta ) + \\
 & c_{\lambda \mu} \frac{\partial \mathcal{F}_\mu}{\partial \nu_\lambda} +
 c_{\mu \eta} \frac{\partial \mathcal{F}_\mu}{\partial \nu_\eta} 
 - 2  c_{\lambda \eta}
\end{split}
\right.
\end{equation}

with:
\begin{equation}
A_{\lambda \eta} =  
\left\{
\begin{split}
\frac{\mathcal{F}_\lambda \, (1/T - \mathcal{F}_\lambda)}{N_\lambda} 
\qquad & \textrm{if  } \lambda=\eta \\
0 \qquad & \textrm{otherwise}
\end{split}
\right.
\end{equation}

Note that, for the concision of the expressions, we used Einstein's
index summation convention: if an index is repeated in a product, a
summation over the whole range of value is implied (e.g. we sum over
\(\lambda \in \{e,i\} \) in the first equation, note that,
consequently, \(\lambda\) does not appear in the left side of the
equation). Also the dependency of the firing rate response to the
excitatory and inhibitory activities has been omitted: yielding
\(\mathcal{F}_\mu\) instead of \(\mathcal{F}_\mu(\nu_e,\nu_i)\),
\(\forall \mu \in \{e,i\} \).

We will also use the reduction to first order of this system (for the
phase-space analysis, see Results). This yields:

\begin{equation}
\label{eq:master-eq-1st-order}
T \, \frac{\partial \nu_\mu}{\partial t} = \mathcal{F}_\mu - \nu_\mu \\
\end{equation}

\subsection{Afferent stimulation}
\label{sec-3-8}

In some simulations, an afferent input was present and was represented
by the following piecewise double Gaussian waveform:

\begin{equation}
\label{eq:input}
\nu_e^{aff}(t) = A \, \Big(
e^{-(\frac{t-t_0}{\sqrt{2} \tau_1})^2} \mathcal{H}(t_0-t)+
e^{-(\frac{t-t_0}{\sqrt{2} \tau_2})^2} \mathcal{H}(t-t_0)
\Big)
\end{equation}

In this afferent input, we can independently control: 1) the maximum
amplitude \(A\) of the stimulation, its rising time constant
\(\tau_1\) and its decay time constant \(\tau_2\).

\section{Results}
\label{sec-4}
\normalsize

The results are organized as follows. We construct the analytical
model that describes the dynamics of a single cortical column. We
start by describing the semi-analytical workflow that enables the
derivation of the cellular transfer function: the core of this
population model. Next, we investigate whether the analytical
description accurately describe population dynamics by comparing its
prediction to numerical simulations. Finally, we investigate the
response of the network model subject to an external input.

\subsection{Modeling a single cortical column}
\label{sec-4-1}

\begin{figure*}
\centering
\includegraphics[width=.7\linewidth]{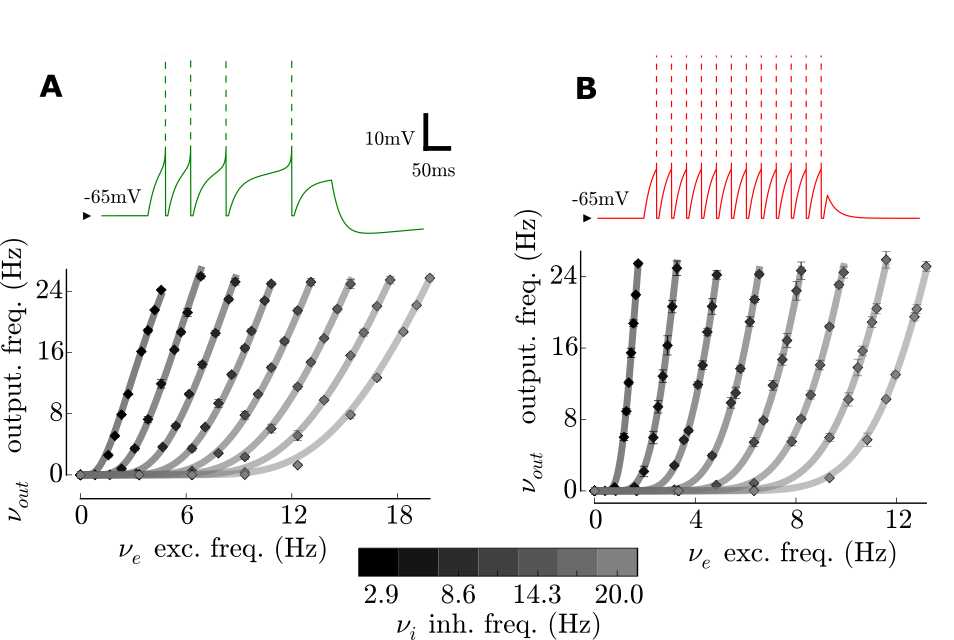}
\caption{\label{fig:tf}\textbf{Single cell models of the excitatory and inhibitory populations.} Top: response to a current step of 200pA lasting 300ms. Bottom: \emph{transfer function} of the single cell, i.e. output firing rate as a function of the excitatory (x-axis) and inhibitory (color-coded) presynaptic release frequencies. Note that the range of the excitatory and frequencies assumes numbers of synapses (\(K_e\)=40 and \(K_i\)=10 for the excitation and inhibition respectively).  \textbf{(A)} Excitatory cells. Note the presence of spike-frequency adaptation and subthreshold adaptation. \textbf{(B)} Inhibitory cells. Note the very narrow spike initiation dynamics (\(k_a\)=0.5mV). Also, note the steepest relation to excitation (with respect to the excitatory cell) at various inhibitory levels as a result of the increased excitability as a result of the increased excitability of the inhibitory cell (with respect to the excitatory cell).}
\end{figure*}

Because optical imaging presumably sample most of its signals from
superficial layers, we model here the layer II/III network: it is
characterized by a strong recurrent connectivity and an important
cellular diversity, in particular one finds many types of interneurons
\cite{Markram2004,Ascoli2008a}. We adopt here a very simplistic
description of this network, it is made of two neuronal population:
one excitatory and one inhibitory comprising 8000 and 2000 neurons
respectively. All neurons within the two population synaptically
interconnect randomly to each other with a connectivity probability of
5\%. The excitatory and inhibitory cells have the same passive
properties. We nonetheless include an asymmetry between the excitatory
and inhibitory populations: because the inhibitory population includes
Fast-Spiking cells that can exhibit very high firing frequencies
\cite{Markram2004}, we set its spiking mechanism sharper (more precisely
its sodium activation activation curve is steeper, see Methods) than
that of excitatory cells, additionally we add a strong spike-frequency
adaptation current in excitatory cells that is absent in inhibitory
cells. Those two effects render the inhibitory neurons more excitable
(see the different responses to the same current step in Figure
\ref{fig:tf}). All parameters of the cortical column can be found in Table
\ref{table:params}.

\subsection{A Markovian model to describe population dynamics}
\label{sec-4-2}

We now want to have an analytical description of the collective
dynamics of this local network. We adopted the formalism presented in
\citetext{ElBoustani2009}. Two reasons motivated this choice: 1) because
10000 neurons is still far from the large network limit, finite-size
effects could have a significant impact on the dynamics and 2) because
of the relative complexity of the cellular models, an analytic
treatment of the type \citetext{Amit1997} is, to our knowledge, not
accessible and would be extremely challenging to derive. The Markovian
framework proposed in \citetext{ElBoustani2009} positively respond to
those two constraints: it is a second-order description of population
activity that describes fluctuations emerging from finite-size effects
and it is applicable to any neuron model as long as its transfer
function can be characterized. In a companion study \cite{Zerlaut2016},
we developed a semi-analytical approach to characterize those
transfer functions (see next section), we will therefore incorporate
this description into the formalism.

Nonetheless, the study of \citetext{ElBoustani2009} only investigated
the ability of the formalism to describe 1) the stationary point of
the network activity and 2) in a situation where the neuronal models
models had an analytic estimate for the transfer function
(current-based integrate-and-fire model). Investigating whether this
description generalizes to transient dynamics and transfer functions
estimated with a semi-analytical approach is investigated in the next
sections.

\subsection{Transfer functions of excitatory and inhibitory cells}
\label{sec-4-3}

We briefly describe here the semi-analytical approach used to
characterize the transfer function (see details in the Methods).

The transfer function \(\mathcal{F}\) of a single neuron is defined
here as the function that maps the value of the stationary excitatory
and inhibitory presynaptic release frequencies to the output
stationary firing rate response, i.e. \(\nu_{out} = \mathcal{F}(\nu_e,
\nu_i)\). This kind of input-output functions lie at the core of
\emph{mean-field} models of population dynamics, reviewed in
\citetext{Renart2004} and is consequently the main ingredient of the
formalism adopted here \cite{ElBoustani2009}. Note here that the
formulation of the transfer function imply a stationary hypothesis:
both for the input (stationary Poisson processes) and the output
firing (a stationary firing rate). We will study in the following what
are the limitations introduced by this stationary hypothesis in the
description of the temporal dynamics of network activity.

In a previous communication \cite{Zerlaut2016}, we found that the firing
rate response of several models (including the adaptative exponential
integrate and fire considered in this study) would be captured by a
\emph{fluctuations-dependent} threshold in a simple approximation of the
firing probability (see Methods).

The semi-analytical approach thus consisted in making numerical
simulations of single-cell dynamics for various presynaptic activity
levels (i.e. scanning various \(\nu_e, \nu_i\) configurations) and
measuring the output firing rate \(\nu_{out}\). All those
configurations corresponded to analytical estimates of
\((\mu_V,\sigma_V,\tau_V)\), we then fitted the
\emph{fluctuations-dependent} threshold that bring the analytical estimate
to the measured firing response. This procedure resulted in the
analytical estimates shown in Figure \ref{fig:tf} and compared with the
results of numerical simulations.

\subsection{Spontaneous activity in the cortical column}
\label{sec-4-4}

\begin{figure*}
\centering
\includegraphics[width=.7\linewidth]{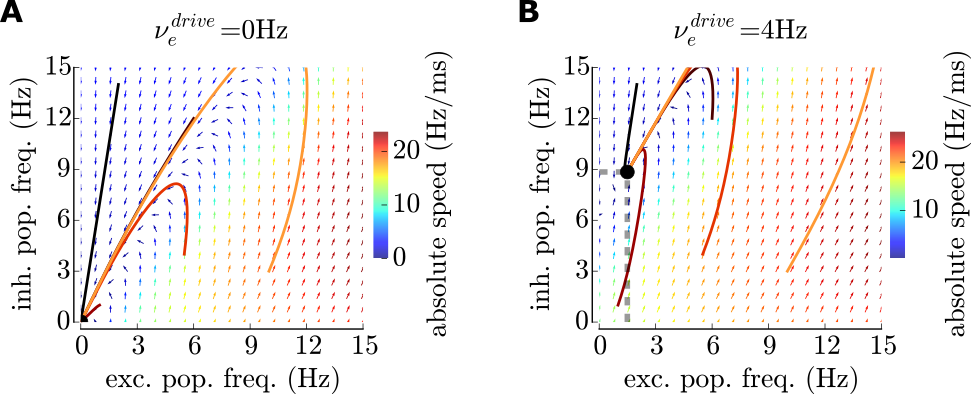}
\caption{\label{fig:phase-space}\textbf{Using the analytical description to look for a stable configuration of spontaneous network activity.} Phase space of the dynamical system resulting from the first order of the markovian description, shown for two levels of external excitatory drive \(\nu_e^{drive}\). The lines represents trajectories resulting from different initial conditions. The vector field correspond to the time-evolution operator (the arrows represent the direction in the two-dimensional space and the color codes for the norm of the vector). \textbf{(A)} Phase space in the absence of an external drive \(\nu_e^{drive}\)=0Hz, the stable fixed point of the dynamics correspond to the quiescent network state \(\nu_e=\nu_i\)=0Hz. \textbf{(A)} Phase space with an external drive \(\nu_e^{drive}\)=4Hz, the stable fixed point of the dynamics correspond now corresponds to an active state with asymmetric activity levels: \(\nu_e\)=1.6Hz and \(\nu_i\)=8.9Hz (round marker).}
\end{figure*}

\begin{figure}[tb!]
\centering
\includegraphics[width=.4\linewidth]{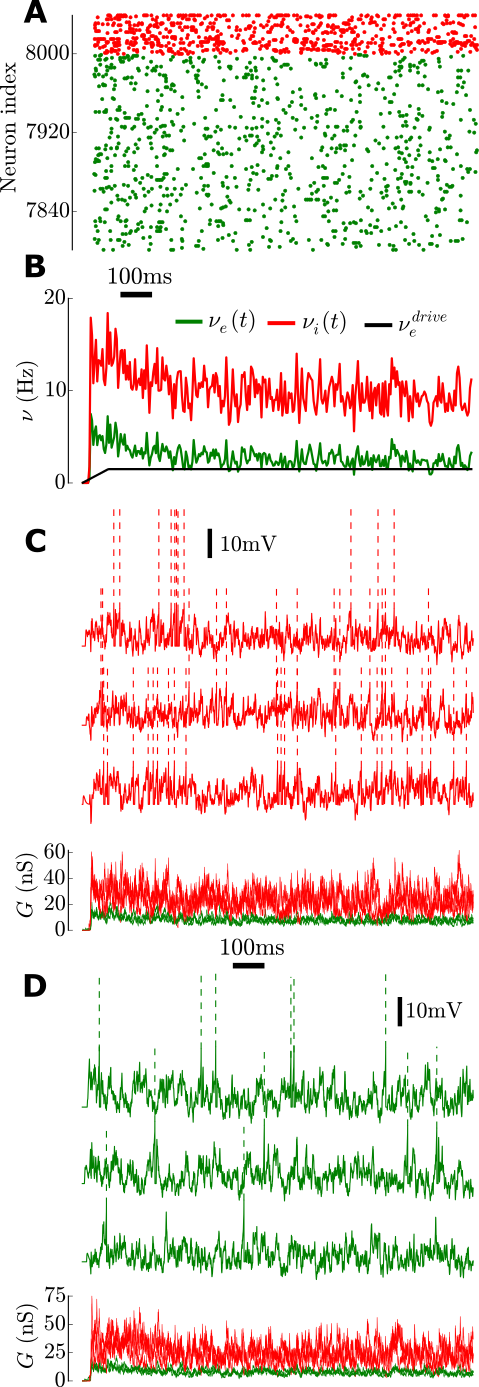}
\caption{\label{fig:ntwk-act}\textbf{Numerical simulations of the dynamics of a recurrent network of 10000 neurons (see parameters in Table \ref{table:params})}. Note that all plots have the same x-axis: time. \textbf{(A)} Sample of the spiking activity of 500 neurons (green, 400 excitatory and red, 100 inhibitory). \textbf{(B)} Population activity (i.e. spiking activity sampled in 5ms time bins across the population) of the excitatory (green) and inhibitory (red) sub-populations. We also show the applied external drive (\(\nu_e^{drive}(t)\), black line), note the slow linear increase to reach \(\nu_e^{drive}\)=4Hz and try to reduce the initial synchronization that would result from an abrupt onset. \textbf{(C)} Membrane potential (top) and conductances (bottom, excitatory in green and inhibitory in red) time courses of three randomly chosen inhibitory neurons. \textbf{(D)} Membrane potential and conductances time courses of three randomly chosen excitatory neurons.}
\end{figure}

\begin{figure}[tb!]
\centering
\includegraphics[width=.8\linewidth]{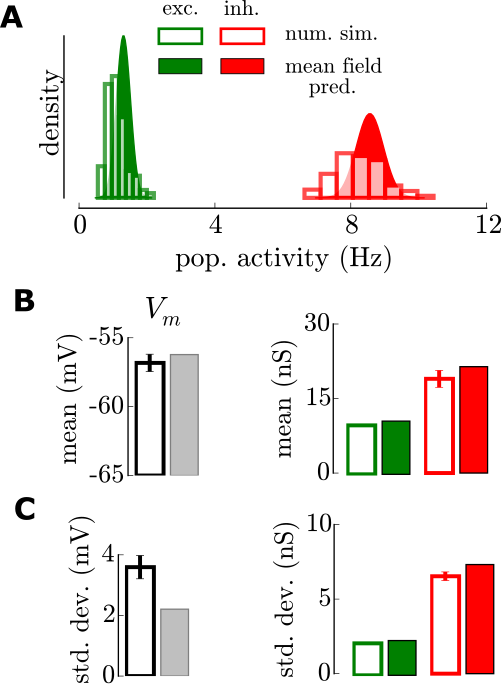}
\caption{\label{fig:mf-stat-pred}\textbf{Mean field prediction of the stationary activity.} Those quantities are evaluated after discarding the initial 500ms transient. \textbf{(A)} Gaussian predictions of the population activities (filled curve) compared to those observed in numerical simulations (empty bars). \textbf{(B)} Mean of the membrane potential and conductances time courses. Evaluated over 3 cells for the numerical simulations (empty bars, mean and standard deviation). \textbf{(C)} Standard deviation of membrane potential and conductances time courses.}
\end{figure}

The combination of the transfer function and the markovian formalism
(Equation \ref{eq:master-equation} in the Methods) yields our analytical
description of the layer II-III population dynamics in a single
cortical column.

We first use this analytical description to look for a physiological
configuration of spontaneous activity. There exists two qualitatively
different types of spontaneous asynchronous activity
\cite{Vogels2005,Kumar2008}: either the network is dominated by
inhibition and the network needs an asynchronous external excitatory
drive to exhibit spontaneous activity \cite{Amit1997,Brunel2000} or
the network exhibits an asynchronous self-sustained activity state and
just needs an initial "kick" to exit from the quiescent state
\cite{Vogels2005,Kumar2008,ElBoustani2009}. In the latter case, the
network is globally dominated by excitation and a strong
\emph{shunting conductance effects} prevents the network from an
excitatory runaway \cite{Kuhn2004,Kumar2008}. Those two behaviors are
thus determined by the membrane, synaptic and connectivity
parameters. We therefore investigate how the chosen network parameters
in this study would determine the qualitative nature of the
spontaneous activity state.

In the case of a single electrophysiological type (e.g. excitatory and
inhibitory neurons taken as the same integrate-and-fire model), it was
shown that a simple \emph{mean-field} analysis allow to predict in which
situation the network parameters corresponds
\cite{Brunel2000,Kumar2008}, here we generalized this approach to the
two populations considered in this study and we investigate the
behavior of our network model given the parameters of Table
\ref{table:params}. To this purpose, we simplified the dynamical system
describing population activity (Equation \ref{eq:master-equation}) to its
first order so that we get a two dimensional system describing the
population spiking activity \(\nu_e(t)\) and \(\nu_i(t)\). We then
plotted the vector field of the time evolution operator in the phase
space of the dynamical system direction and launched some trajectories
with different initial conditions (see Figure \ref{fig:phase-space}). The
result of this analysis is that, in absence of external input
(\(\nu_e^{drive}\)=0Hz), the only fixed point of the system is the
quiescent state (see Figure \ref{fig:phase-space}A). This prediction of the
mean-field analysis was indeed confirmed by numerical simulations,
whatever the initial external "kick", the activity rapidly decayed
(T<50ms) to the quiescent state.

We conclude that, given the parameters of Table \ref{table:params}, our
network model does not have the ability to self-sustain activity and
will need an external excitatory drive to exhibit spontaneous activity
(note that this is also consistently with recent \emph{in vivo}
observations in mice visual cortex, see
\citetext{Reinhold2015}). Indeed, when raising the external drive, a
non-quiescent fixed point appears (see Figure \ref{fig:phase-space}B for
\(\nu_e^{drive}\)=4Hz). Numerical simulations confirmed the existence
of such a fixed point at those levels of activity (see Figure
\ref{fig:ntwk-act}).

The particularity of this stationary fixed-point is its asymmetry in
terms of population activity, it corresponds to \(\nu_e\)=1.6Hz and
\(\nu_i\)=8.9Hz (i.e. corresponding to a factor 5-6 between the their
respective firing rates). The origin of this asymmetry is very
naturally the asymmetry in electrophysiological properties as the
excitatory and inhibitory neurons sample statistically the same
recurrent and external input. This phenomena has been observed in
extracellular recordings in human cortex \cite{Peyrache2012}, cells
categorized as Fast-Spiking (such as our inhibitory cells) were shown
to fire 6-7 times more than cells categorized as Regular-Spiking (such
as our excitatory cells), an asymmetry in excitabilities thus
naturally provides a putative explanation for this phenomena (rather
than specific circuitry).

\subsection{Accuracy of the description of the spontaneous activity state}
\label{sec-4-5}

We compare more closely the numerical simulation (Figure
\ref{fig:ntwk-act}) to the prediction of the Markovian description.

First, we see that there is a transient period of \(\sim\) 400ms
resulting from the onset of the external drive (see Figure
\ref{fig:ntwk-act}B-D), we will therefore evaluate stationary properties
after discarding the first 500ms of the simulation.

After this initial transient, the population activities (\(\nu_e\) and
\(\nu_i\)) fluctuates around the stationary levels (see Figure
\ref{fig:ntwk-act}). The Markovian description predicts this phenomena as it
contains the impact of finite size effects (the network comprises
10000 neurons). In Figure \ref{fig:mf-stat-pred}A, we can see that the
distributions of the excitatory and inhibitory population activities
are rather well predicted by the formalism (it slightly overestimates
the means of the population activities).

We also investigated whether the average neuronal and synaptic
quantities were well predicted by the Markovian formalism. Indeed, we
found a very good match for all quantities (see Figure
\ref{fig:mf-stat-pred}B,C, mean and variance of membrane potential and
synaptic conductances). Only the standard deviation of the membrane
potential fluctuations was underestimated (Figure \ref{fig:mf-stat-pred}C),
presumably because of residual synchrony in the dynamics whereas the
Markovian formalism assumes a purely asynchronous regime.

\subsection{Description of the response to time-varying input}
\label{sec-4-6}

\begin{figure}[tb!]
\centering
\includegraphics[width=.4\linewidth]{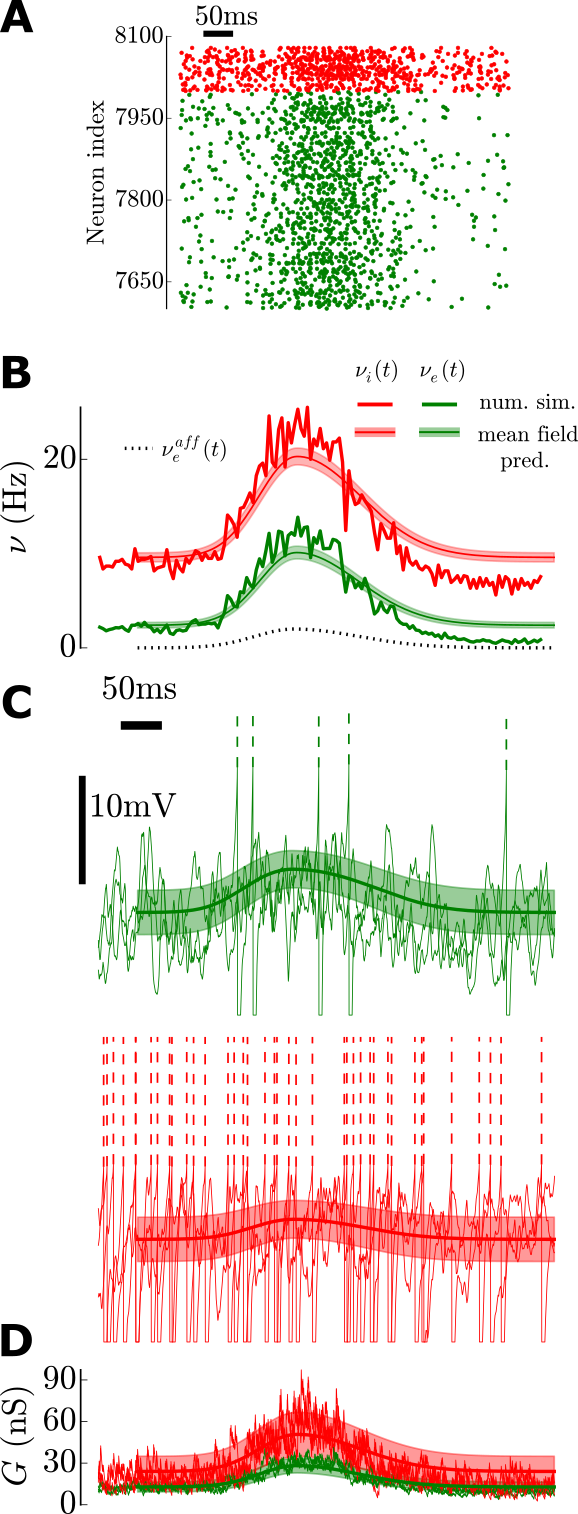}
\caption{\label{fig:mf-temp-pred}\textbf{Network response to a time-varying input and associated prediction of the Markovian formalism.} For all plots, the x-axis corresponds to time. Shown after 500ms of initial stimulation. \textbf{(A)} Sample of the spiking activity of 500 neurons (green, 400 excitatory and red, 100 inhibitory). \textbf{(B)} Population activity (in 5ms bins) of the excitatory (green) and inhibitory (red) sub-populations. Superimposed is the mean and standard deviation over time predicted by the Markovian formalism. We also show the applied external stimulation (\(\nu_e^{aff}(t)\), dotted line). \textbf{(C)} Membrane potential time courses of three excitatory cells (green, top) and three inhibitory cells (red, bottom) with the prediction of the mean and standard deviation in time. \textbf{(D)} Conductance time courses of the six cells in \textbf{C} with the predictions of the fluctuations superimposed.}
\end{figure}

We now examine whether the formalism captures the response to
time-varying input. Here again, we set the input and examine the
response after 500ms of initial simulation to discard transient
effects.

We first choose an afferent input of relatively low frequency content
(\(\sim\) [5-20]Hz, \( \tau_1 \)=60ms and \( \tau_2 \)=100ms in
Equation \ref{eq:input}). The afferent input waveform, formulated in terms
of firing rate, was translated into individual afferent spikes
targeting the excitatory population. The response of the network to
this input is shown in Figure \ref{fig:mf-temp-pred} in comparison with the
prediction of the Markovian formalism. The excitatory population
activity raises and immediately entrains a raise of the inhibitory
population. The analytical description captures well the order of
magnitude of the deflection, it only slightly underestimates the peak
value (Figure \ref{fig:mf-temp-pred}B). But the numerical simulations also
show a marked hyperpolarization after the stimulation, the return to
the baseline level happens only \(\sim\) 200-300 ms after the end of
the stimulus, and not immediately as predicted by the Markovian
framework. Here this strong hyperpolarization is the result of the
strong spike-frequency adaptation current that remains as a
consequence of the high activity evoked by the stimulus. In the
Markovian there is no memory of the previous activity and therefore
this phenomena can not be accounted for. This typically illustrates a
limitation of the analytical description provided here. Note that this
is not a fundamental limitation of the Markovian formalism, it is a
limitation of this version of the formalism, that contains only
variables related to the instantaneous activity (see Discussion).

\begin{figure}[tb!]
\centering
\includegraphics[width=.7\linewidth]{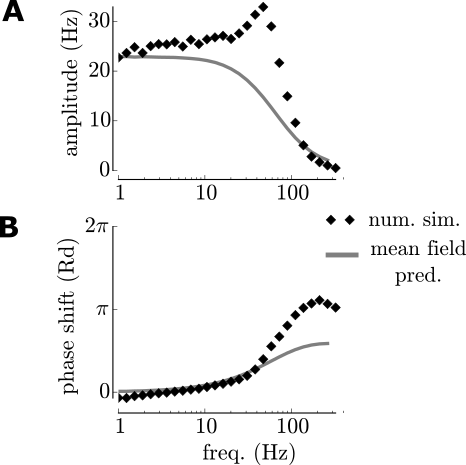}
\caption{\label{fig:mf-freq-dep}\textbf{Limitations of the Markovian description in the frequency domain.} Response of the network (numerical simulation and analytical description) to sinusoidal stimulation of the form \(\nu_e^{aff} = 5\mathrm{Hz} \, \big(1-\cos(2 \pi f (t-t_0))\big)/2\). The stimulation was set on at \(t_0\)=500ms. The response was fitted by a function of the form  \(\nu(t) = A \, \big(1-\cos(2 \pi f (t-t_0) - \phi)\big)/2\). \textbf{(A)} Amplitude of the sinusoidal response (\(A\) in the fitted response) for various frequencies. \textbf{(B)} Phase shift of the sinusoidal response (\(\phi\) in the fitted response) for various frequencies.}
\end{figure}

To study more precisely the temporal validity of the formalism, we
modulated the network activity by sinusoidal input and compared the
response predicted by the analytical description.

The numerical simulations showed a marked resonance at
\(\sim\)50Hz. Given the relatively high strength (compared to the
external input) of the excitatory-inhibitory loop, the network is
close to a bifurcation toward oscillations that are typically in the
gamma range \cite{Brunel2003}. A sinusoidal input therefore amplifies
those frequencies \cite{Ledoux2011}. Because the individual excitatory
and inhibitory post-synaptic currents approximately match each other,
the theoretical study of \citetext{Brunel2003} would predict
oscillations at 50-60Hz (the bifurcation would be achieved by reducing
\(\tau_e\)), thus compatible with the present observation.

More importantly, the main insight of this analysis is to show that
the network can track very fast temporal variations in the input, even
at time scales smaller than the integration time constant of the
single neurons \cite{VanVreeswijk1996}.  Recurrent neural networks
globally behave as low-pass filters (though see \citetext{Ledoux2011} for a
detailed treatment of the appearance of resonances), but with a high
cutoff frequency compared to the frequency content of thalamic input
for classical artificial stimuli (e.g. in the visual system: drifting
gratings, supra-10ms flashes, etc\ldots{}).

Leaving apart the failure of capturing the network resonance (that is
linked to this special configuration of synaptic parameters), we
conclude that in the frequency range that will be used in the
following (f<50-100Hz) the description of the formalism gives a
relatively accurate description of the network response in the sense
that it accurately predicts that there should not be a frequency
filtering within this range. Again, \emph{in vivo} experiments in awake
mice suggested that V1 cortical networks had a cut-off frequency above
this range (\(\sim\)100Hz in \citetext{Reinhold2015}).

Thus, by comparing numerical simulations of network dynamics and the
Markovian formalism, we showed that, despite some discrepancies, this
analytical framework describes both the spontaneous activity and the
response in the [0,100]Hz range of a sparsely connected recurrent
network of distinct excitatory and inhibitory cells.

\section{Discussion}
\label{sec-5}
\normalsize

In the present study, we investigated {a mean-field model of
  networks with different electrophysiological properties, described
  using the AdEx model with conductance-based synapses.  We found
  that} the Markovian formalism proposed in \citetext{ElBoustani2009}
was able to describe the {steady-state and} temporal dynamics of
{such} networks. Though this formalism was shown to be a
relatively accurate description of the response simulated in numerical
networks, we also showed the limits of this formalism. The relative
complexity of the theoretical problem should be stressed: our model
includes non-linear phenomena such as an voltage-dependent activation
curve for spike emission or spike frequency. The proposed
semi-analytical appraoch thus offers a convenient description for
theoretical models where an exact analytical treatment would not be
achievable.

Unlike previous studies
\cite{Brunel2000,Vogels2005,Kumar2008,ElBoustani2009}, we considered
networks of non-linear integrate-and-fire neurons with asymmetric
electrophysiological properties between excitatory and inhibitory
cells. This type of network is more realistic because it includes the
adaptation properties of excitatory cells, and the fact that
inhibitory cells are more excitable and fire at higher rates.  We
could demonstrate the relative accuracy of the markovian formalism
(with the semi-analytical approach) in a situation including this
increased complexity. The mean-field model obtained was able to
predict the level of spontaneous activity of the network, as well as
its response to external time-varying inputs.

This versatile theoretical description of the local cortical network
could be improved. For example the strong hyperpolarization of
population activity after a transient rise (see Figure
\ref{fig:mf-temp-pred}B) was shown to be missed by the mean-field
formalism. Indeed, this version does not have a memory of the previous
activity levels and thus can not account for the effect of the
long-lasting spike-frequency adaptation mechanism that has been
strongly activated by the activity evoked by the stimulus. One could
design another version of the Markovian formalism to capture such
adaptation-mediated effects. Instead of accounting for adaptation
within the \emph{transfer function} (i.e. accounting only for its
stationary effects), one can introduce a new variable with a
dependency on time and activity: a ``population adaptation current'',
that can directly be derived from the equation of the AdExp model.
Investigating the accuracy of such theoretical descriptions should be
the focus of future work. 

{We conclude by proposing the present model as a good candidate
  for modeling VSDi data.  Not only the present mean-field framework
  gives access to the mean voltage and its time evolution, but it
  could easily be extended to model VSDi signals.  The present model
  represents a local population of cortical excitatory and inhibitory
  neurons, and thus can be thought to represent a ``pixel'' of the
  VSDi.  The full VSDi model could be obtained by embedding the
  present local population description within a spatial model, and
  yield 1-D or 2-D ensembles of such pixels, and thereby model VSDi
  recordings in mammalian neocortex.  This is the focus of current
  investigations \cite{Destexhe2015}.}

\subsection*{Acknowledgments} 

Research supported by the CNRS, the ICODE excellence network, and the
European Community (Human Brain Project, H2020-720270).  Y.Z. was
supported by fellowships from the Initiative d'Excellence Paris-Saclay
and the Fondation pour la Recherche M\'edicale (FDT 20150532751).

\section{References}
\label{sec-6}
\small

\bibliography{biblio}

\begin{thebibliography}{}

\bibitem[\protect\citeauthoryear{Amit and Brunel}{1997}]{Amit1997}
Amit DJ, Brunel N (1997)
\newblock {Model of global spontaneous activity and local structured activity
  during delay periods in the cerebral cortex.}
\newblock {\em Cerebral Cortex}~7:\mbox{237--252}.

\bibitem[\protect\citeauthoryear{Arieli  \bgroup et al.\egroup
  }{1996}]{Arieli1996}
Arieli a, Sterkin a, Grinvald a, Aertsen A, An JH (1996)
\newblock {Dynamics of ongoing activity: explanation of the large variability
  in evoked cortical responses.}
\newblock {\em Science (New York, N.Y.)}~273:\mbox{1868--71}.

\bibitem[\protect\citeauthoryear{Ascoli  \bgroup et al.\egroup
  }{2008}]{Ascoli2008a}
Ascoli GAG, Alonso-Nanclares L, Anderson SA, Barrionuevo G, Benavides-Piccione
  R, Burkhalter A, Buzs{\'{a}}ki G, Cauli B, Defelipe J, Fair{\'{e}}n A, Others
  (2008)
\newblock {Petilla terminology: nomenclature of features of GABAergic
  interneurons of the cerebral cortex}.
\newblock {\em Nature Reviews {\ldots}}~9:\mbox{557--568}.

\bibitem[\protect\citeauthoryear{Berger  \bgroup et al.\egroup
  }{2007}]{Berger2007}
Berger T, Borgdorff A, Crochet S, Neubauer FB, Lefort S, Fauvet B, Ferezou I,
  Carleton A, L{\"{u}}scher HR, Petersen CCH (2007)
\newblock {Combined voltage and calcium epifluorescence imaging in vitro and in
  vivo reveals subthreshold and suprathreshold dynamics of mouse barrel
  cortex.}
\newblock {\em Journal of neurophysiology}~97:\mbox{3751--3762}.

\bibitem[\protect\citeauthoryear{Brette and Gerstner}{2005}]{Brette2005a}
Brette R, Gerstner W (2005)
\newblock {Adaptive exponential integrate-and-fire model as an effective
  description of neuronal activity}.
\newblock {\em Journal of neurophysiology}~\mbox{pp. 3637--3642}.

\bibitem[\protect\citeauthoryear{Brunel}{2000}]{Brunel2000}
Brunel N (2000)
\newblock {Dynamics of sparsely connected networks of excitatory and inhibitory
  spiking neurons}.
\newblock {\em Journal of computational neuroscience}~8:\mbox{183--208}.

\bibitem[\protect\citeauthoryear{Brunel and Hakim}{1999}]{Brunel1999}
Brunel N, Hakim V (1999)
\newblock {Fast global oscillations in networks of integrate-and-fire neurons
  with low firing rates}.
\newblock {\em Neural computation}~11:\mbox{1621--1671}.

\bibitem[\protect\citeauthoryear{Brunel and Wang}{2003}]{Brunel2003}
Brunel N, Wang XJ (2003)
\newblock {What determines the frequency of fast network oscillations with
  irregular neural discharges? I. Synaptic dynamics and excitation-inhibition
  balance.}
\newblock {\em Journal of neurophysiology}~90:\mbox{415--430}.

\bibitem[\protect\citeauthoryear{Chemla and Chavane}{2010}]{Chemla2010}
Chemla S, Chavane F (2010)
\newblock {A biophysical cortical column model to study the multi-component
  origin of the VSDI signal}.
\newblock {\em NeuroImage}~53:\mbox{420--438}.

\bibitem[\protect\citeauthoryear{Civillico and Contreras}{2012}]{Civillico2012}
Civillico EF, Contreras D (2012)
\newblock {Spatiotemporal properties of sensory responses in vivo are strongly
  dependent on network context.}
\newblock {\em Frontiers in systems neuroscience}~6:\mbox{25}.

\bibitem[\protect\citeauthoryear{Contreras and Llinas}{2001}]{Contreras2001}
Contreras D, Llinas R (2001)
\newblock {Voltage-sensitive dye imaging of neocortical spatiotemporal dynamics
  to afferent activation frequency.}
\newblock {\em The Journal of neuroscience : the official journal of the
  Society for Neuroscience}~21:\mbox{9403--9413}.

\bibitem[\protect\citeauthoryear{Daley and Vere-Jones}{2007}]{Daley2007}
Daley DJ, Vere-Jones D (2007)
\newblock {\em An introduction to the theory of point processes: volume II:
  general theory and structure}, Vol.~2
\newblock Springer Science \& Business Media.

\bibitem[\protect\citeauthoryear{Destexhe \bgroup et al.\egroup
  }{2003}]{Destexhe2003}
Destexhe A, Rudolph M, Par\'{e} D (2003)
\newblock {The high-conductance state of neocortical neurons in vivo}.
\newblock {\em Nature Reviews Neuroscience}~4:\mbox{739--751}.

\bibitem[\protect\citeauthoryear{Destexhe  \bgroup et al.\egroup
  }{2015}]{Destexhe2015}
Destexhe A, Zerlaut Y, Reynaud A, Chemla S, Chavane F (2015)
\newblock {Conductance-Based Interactions Predict The Suppressive Effect Of
  Interacting Propagating Waves In Awake Monkey Visual Cortex}.
\newblock {\em Society for Neuroscience, conference abstract}~.

\bibitem[\protect\citeauthoryear{{El Boustani} and
  Destexhe}{2009}]{ElBoustani2009}
{El Boustani} S, Destexhe A (2009)
\newblock {A master equation formalism for macroscopic modeling of asynchronous
  irregular activity states}.
\newblock {\em Neural computation}~21:\mbox{46--100}.

\bibitem[\protect\citeauthoryear{Ferezou \bgroup et al.\egroup
  }{2006}]{Ferezou2006}
Ferezou I, Bolea S, Petersen CCH (2006)
\newblock {Visualizing the Cortical Representation of Whisker Touch:
  Voltage-Sensitive Dye Imaging in Freely Moving Mice}.
\newblock {\em Neuron}~50:\mbox{617--629}.

\bibitem[\protect\citeauthoryear{Goodman and Brette}{2009}]{Goodman2009}
Goodman DFM, Brette R (2009)
\newblock {The brian simulator}.
\newblock {\em Frontiers in Neuroscience}~3:\mbox{192--197}.

\bibitem[\protect\citeauthoryear{Kuhn \bgroup et al.\egroup }{2004}]{Kuhn2004}
Kuhn A, Aertsen A, Rotter S (2004)
\newblock {Neuronal integration of synaptic input in the fluctuation-driven
  regime.}
\newblock {\em The Journal of neuroscience : the official journal of the
  Society for Neuroscience}~24:\mbox{2345--56}.

\bibitem[\protect\citeauthoryear{Kumar  \bgroup et al.\egroup
  }{2008}]{Kumar2008}
Kumar A, Schrader S, Aertsen A, Rotter S (2008)
\newblock {The high-conductance state of cortical networks}.
\newblock {\em Neural Computation}~20:\mbox{1--43}.

\bibitem[\protect\citeauthoryear{Latham  \bgroup et al.\egroup
  }{2000}]{Latham2000}
Latham PE, Richmond BJ, Nelson PG, Nirenberg S (2000)
\newblock {Intrinsic Dynamics in Neuronal Networks. I. Theory}.
\newblock {\em J Neurophysiol}~83:\mbox{808--827}.

\bibitem[\protect\citeauthoryear{Ledoux and Brunel}{2011}]{Ledoux2011}
Ledoux E, Brunel N (2011)
\newblock {Dynamics of networks of excitatory and inhibitory neurons in
  response to time-dependent inputs.}
\newblock {\em Frontiers in computational neuroscience}~5:\mbox{25}.

\bibitem[\protect\citeauthoryear{Markram  \bgroup et al.\egroup
  }{2015}]{Markram2015}
Markram H, Muller E, Ramaswamy S, Reimann Mea (2015)
\newblock {Reconstruction and Simulation of Neocortical Microcircuitry}.
\newblock {\em Cell}~163:\mbox{456--492}.

\bibitem[\protect\citeauthoryear{Markram  \bgroup et al.\egroup
  }{2004}]{Markram2004}
Markram H, Toledo-Rodriguez M, Wang Y, Gupta A, Silberberg G, Wu C (2004)
\newblock {Interneurons of the neocortical inhibitory system.}
\newblock {\em Nature reviews. Neuroscience}~5:\mbox{793--807}.

\bibitem[\protect\citeauthoryear{McCormick  \bgroup et al.\egroup
  }{1985}]{McCormick1985}
McCormick DA, Connors BW, Lighthall JW, Prince Da (1985)
\newblock {Comparative electrophysiology of pyramidal and sparsely spiny
  stellate neurons of the neocortex.}
\newblock {\em Journal of neurophysiology}~54:\mbox{782--806}.

\bibitem[\protect\citeauthoryear{Papoulis}{1991}]{Papoulis1991}
Papoulis A (1991)
\newblock {\em Probability, random variables and stochastic processes}
\newblock McGraw-Hill.

\bibitem[\protect\citeauthoryear{Petersen and Sakmann}{2001}]{Petersen2001}
Petersen CCH, Sakmann B (2001)
\newblock {Functionally Independent Columns of Rat Somatosensory Barrel Cortex
  Revealed with Voltage-Sensitive Dye Imaging}.
\newblock {\em The Journal of neuroscience : the official journal of the
  Society for Neuroscience}~21:\mbox{8435--8446}.

\bibitem[\protect\citeauthoryear{Peyrache  \bgroup et al.\egroup
  }{2012}]{Peyrache2012}
Peyrache A, Dehghani N, Eskandar EN, Madsen JR, Anderson WS, Donoghue Ja,
  Hochberg LR, Halgren E, Cash SS, Destexhe A (2012)
\newblock {Spatiotemporal dynamics of neocortical excitation and inhibition
  during human sleep.}
\newblock {\em Proceedings of the National Academy of Sciences of the United
  States of America}~109:\mbox{1731--6}.

\bibitem[\protect\citeauthoryear{Reinhold \bgroup et al.\egroup
  }{2015}]{Reinhold2015}
Reinhold K, Lien AD, Scanziani M (2015)
\newblock {Distinct recurrent versus afferent dynamics in cortical visual
  processing}.
\newblock {\em Nature Neuroscience}~18.

\bibitem[\protect\citeauthoryear{Renart \bgroup et al.\egroup
  }{2004}]{Renart2004}
Renart A, Brunel N, Wang XJ (2004)
\newblock {Mean-field theory of irregularly spiking neuronal populations and
  working memory in recurrent cortical networks}.
\newblock {\em Computational neuroscience: A comprehensive approach}~\mbox{pp.
  431--490}.

\bibitem[\protect\citeauthoryear{Reynaud \bgroup et al.\egroup
  }{2012}]{Reynaud2012}
Reynaud A, Masson GS, Chavane F (2012)
\newblock {Dynamics of local input normalization result from balanced short-
  and long-range intracortical interactions in area V1.}
\newblock {\em The Journal of neuroscience : the official journal of the
  Society for Neuroscience}~32:\mbox{12558--69}.

\bibitem[\protect\citeauthoryear{Steriade \bgroup et al.\egroup
  }{2001}]{Steriade2001}
Steriade M, Timofeev I, Grenier F (2001)
\newblock {Natural waking and sleep states: a view from inside neocortical
  neurons.}
\newblock {\em Journal of neurophysiology}~85:\mbox{1969--1985}.

\bibitem[\protect\citeauthoryear{van Vreeswijk and
  Sompolinsky}{1996}]{VanVreeswijk1996}
van Vreeswijk C, Sompolinsky H (1996)
\newblock {Chaos in neuronal networks with balanced excitatory and inhibitory
  activity.}
\newblock {\em Science (New York, N.Y.)}~274:\mbox{1724--6}.

\bibitem[\protect\citeauthoryear{Vogels and Abbott}{2005}]{Vogels2005}
Vogels TP, Abbott LF (2005)
\newblock {Signal propagation and logic gating in networks of
  integrate-and-fire neurons}.
\newblock {\em The Journal of neuroscience}~25:\mbox{10786--10795}.

\bibitem[\protect\citeauthoryear{Zerlaut  \bgroup et al.\egroup
  }{2016}]{Zerlaut2016}
Zerlaut Y, Telenczuk B, Deleuze C, Bal T, Ouanounou G, Destexhe A (2016)
\newblock {Heterogeneous firing response of mice layer V pyramidal neurons in
  the fluctuation-driven regime}.
\newblock {\em The Journal of Physiology}~594:\mbox{3791--808}.

\end{thebibliography}

\end{document}